\documentclass[a4paper,fleqn]{cas-sc}

\usepackage{xcolor}
\definecolor{ra}{rgb}{0.8, 0.0, 0.0}
 
\usepackage[numbers,sort&compress]{natbib}
\usepackage{bm}
\usepackage{cite}
\def\tsc#1{\csdef{#1}{\textsc{\lowercase{#1}}\xspace}}
\tsc{WGM}
\tsc{QE}
\tsc{EP}
\tsc{PMS}
\tsc{BEC}
\tsc{DE}

\begin{document}
\let\WriteBookmarks\relax
\def\floatpagepagefraction{1}
\def\textpagefraction{.001}
\shorttitle{Modeling financial time series with $\phi^{4}$ quantum field theory}
\shortauthors{D. Bachtis et~al.}

\title [mode = title]{Modeling financial time series with $\phi^{4}$ quantum field theory}                      



\author[1]{Dimitrios Bachtis}
\cormark[1]
\ead{d.bachtis@qmul.ac.uk}

\affiliation[1]{organization={Centre for Theoretical Physics, Department of Physics and Astronomy,
Queen Mary University of London},
                city={London},
                postcode={E1 4NS}, 
                country={United Kingdom}}
{}
\author[1]{David S. Berman}
\ead{d.s.berman@qmul.ac.uk}
\author[1]{Arabella Schelpe}
\ead{cacs2@cantab.net}



\begin{abstract}
We use a $\phi^{4}$ quantum field theory with inhomogeneous couplings and explicit symmetry-breaking to model an ensemble of financial time series from the S$\&$P 500 index. The continuum nature of the $\phi^4$ theory avoids the inaccuracies that occur in Ising-based models which require a discretization of the time series. We demonstrate this using the example of the 2008 global financial crisis. The $\phi^{4}$ quantum field theory is expressive enough to reproduce the higher-order statistics such as the market kurtosis, which can serve as an indicator of possible market shocks. Accurate reproduction of high kurtosis is absent in binarized models. Therefore Ising models, despite being widely employed in econophysics, are incapable of fully representing empirical financial data, a limitation not present in the generalization of the $\phi^{4}$ scalar field theory. We then investigate the scaling properties of the $\phi^{4}$ machine learning algorithm and extract exponents which govern the behavior of the learned couplings (or weights and biases in ML language) in relation to the number of stocks in the model.  Finally, we use our model to forecast the price changes of the AAPL, MSFT, and NVDA stocks. We conclude by discussing how the $\phi^{4}$ scalar field theory could be used to build investment strategies and the possible intuitions that the quantum field theoretical operations of dimensional compactification and renormalization can provide for financial modelling.
\end{abstract}

%
%

%


\maketitle

\section{\label{sec:level1} Introduction}

 Most financial market data is represented by time series, for example describing the evolution of stock prices. The study of financial time series allows one to construct trading, portfolio optimization and risk management strategies, and potentially early warning indicators for major financial events, such as global crises. Time series have been historically studied with various methods, such as linear and nonlinear~\citep{hsieh1,Potters1}, heteroscedastic~\citep{Engle1982,BOLLERSLEV1986307,Nelson1991,BOLLERSLEV19925}, or continuous-time ~\citep{fisch,kou1,ait1} models, principal component analysis and factor models~\citep{Chen1,bai1,connor1}, state-space models and Kalman filters~\citep{siem1,kalman1960new}, Markov chain Monte Carlo simulations~\citep{dempster1,sangj1,CHIB2002281,jacq1}, log periodic power law~\citep{johansen2000crashes,SORNETTE2006153}, time-averaged~\citep{Cherstvy_2017,Ritschel_2021} methods, and machine learning~\citep{10.1098/rsta.2020.0209}. 


A physical system commonly employed to study aspects of finance is the Ising model due to its $Z_{2}$ symmetry, second-order phase transition, and the binary degrees of freedom which enable a representation of buy and sell dynamics~\citep{PhysRevLett.89.158701,sornzhou,Chowdhury1999}. The Ising model has been employed to study financial time series~\citep{Borysov2015,Zeng_2014,MASKAWA2002563,Bury2013}, but it manifests limitations due to the binary degrees of freedom which necessitate a discretization of empirical financial data. The binarization~\footnote{We mostly use the term binarization in this paper, to refer to the transformation of continuous return values into $(-1,1)$, although it is commonly referred to less precisely as discretization in the literature.} of time series results in loss of crucial information about financial markets, for instance in the reproduction of high-order statistics that can serve as indicators of major financial events.

In this manuscript we propose the $\phi^{4}$ quantum field theory, a system with $Z_{2}$ symmetry and continuous degrees of freedom, to model and forecast empirical financial time series. The $\phi^{4}$ field theory is a generalization of the Ising model, and reduces to the Ising model in a mathematical limit. This suggests that the $\phi^{4}$ theory should be as
 successful as the Ising model, with the additional benefit of being able to model continuous degrees of freedom.
 In addition, the $\phi^{4}$ theory is a nonlinear system, allowing for more complicated dependencies, while retaining complete interpretability. We investigate whether the $\phi^{4}$ theory strikes a desirable balance between
 simplicity and representational capacity.

We specifically explore whether the $\phi^{4}$ quantum field theory can exhibit the advantages of the Ising model, without its limitations.  To our knowledge, the $\phi^{4}$ scalar field theory has never been implemented to study financial time series.  More general $\phi^{n}$ models  have been suggested for portfolio optimization~\citep{SORNETTE200019}, but without an exploration of the specific capabilities of the simpler restriction to $n=4$. The $\phi^{4}$ theory has been additionally proposed as a multi-agent system of financial markets in ~\citep{bachtis2024latticephi4fieldtheory}, while ~\citep{Schmidhuber2020TrendsRA} and ~\citep{Halperin} have also recently worked on similar equations, applying them to trend modelling and multi-asset systems respectively, although neither has made use of the field theory framework.

We initiate our study by considering time series which include events such as the global financial crisis of 2008 in order to investigate if the disordered $\phi^{4}$ model, which is mathematically equivalent to a machine learning algorithm of a Markov random field~\citep{PhysRevD.103.074510}, is able to faithfully reproduce market statistics that vanish in binarized series. 

We then clarify how the $\phi^{4}$ machine learning algorithm is described by a set of interpretable weights and biases, in contrast to algorithms such as neural networks with hidden variables which lead to loss of interpretability of results. Consequently, we investigate scaling properties~\citep{NEURIPS2024_648a5a59} by extracting exponents which describe the behavior of the weights and biases in relation to the number of stocks present within the S$\&$P 500 index.

Finally, we investigate whether the $\phi^{4}$ scalar field theory, which is able to directly model the price changes of stocks, can be utilized to forecast time series. Specifically, we employ techniques of probabilistic machine learning pertinent to the sampling of missing degrees of freedom to predict the price changes for the AAPL, MSFT, and NVDA stocks.

We conclude by discussing how the $\phi^{4}$ scalar field theory could be employed to establish investment strategies and a framework for financial markets developed within quantum field theory based on the mathematical operations of dimensional compactification and renormalization.

\section{\label{sec:level2}  $\phi^{4}$ quantum field-theoretic machine learning}

We define a discretized and fully disordered $\phi^{4}$ theory on a complete graph which is represented by the lattice action:
\begin{equation}\label{eq:orig}
S=-\sum_{i,j}  w_{ij} \phi_{i}\phi_{j}+\sum_{i}\mu_{i} \phi_{i}^{2}+\sum_{i}\lambda_{i}\phi_{i}^{4}-\sum_{i} a_{i}\phi_{i},
\end{equation}
where $w_{ij}$ is an inhomogeneous coupling connecting the $\phi_{i}$ and $\phi_{j}$ fields, $\mu_{i}$ and $\lambda_{i}$ are the inhomogeneous squared mass and lambda couplings, and $a_{i}$ represents an external field that breaks explicitly and locally the $Z_{2}$ symmetry. We denote the set of all lattice sites as $\Lambda$ and the volume of the system as $V$. The $\phi^{4}$ scalar field theory is described by a Boltzmann probability distribution of the form:
\begin{equation}
p(\bm{\phi},\theta)= \frac{\exp[-S(\bm{\phi},\theta)]}{Z},
\end{equation}
where $Z=\int_{-\infty}^{\infty} \exp[-S(\bm{\phi},\theta)] d\bm{\phi}$ is the partition function of the system, $\bm{\phi}$ denotes a configuration comprising the fields $\lbrace \phi_{i} \rbrace$, and  $\theta$ defines the set of the inhomogeneous couplings which are equivalent to learnable variational parameters within a machine learning setting. That is, $\theta$ denotes the full set of learnable couplings $\{ w_{ij},\mu_{i}, \lambda_{i}, a_{i} \} $.  From the perspective of physics, the system discussed herein is simultaneously a $\phi^{4}$ glass~\citep{phi4glass} and a $\phi^{4}$ inhomogeneous external field model.  From the perspective of machine learning, the system discussed herein satisfies the Hammersley-Clifford theorem and is therefore a Markov random field~\citep{PhysRevD.103.074510}. As such the system possesses all the properties advantageous for machine learning of a Markov random field such as efficient gradient-based training and the applicability of simple and scalable sampling methods.

\begin{figure*}[pos=ht]
\center
\includegraphics[width=10cm]{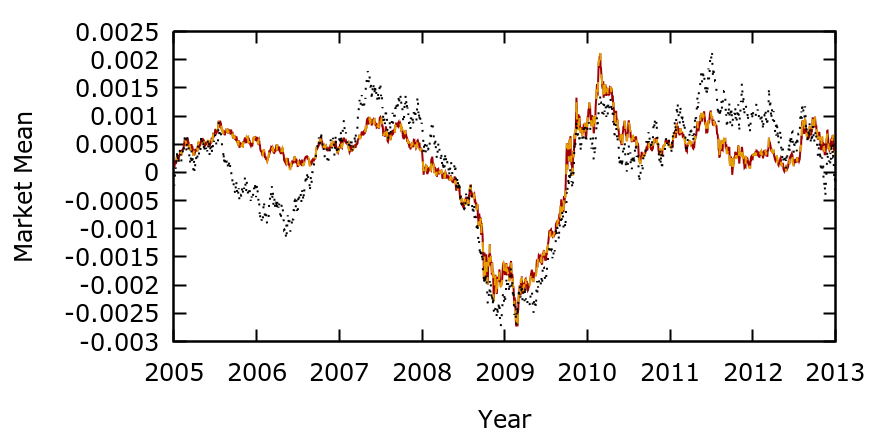}
\includegraphics[width=10cm]{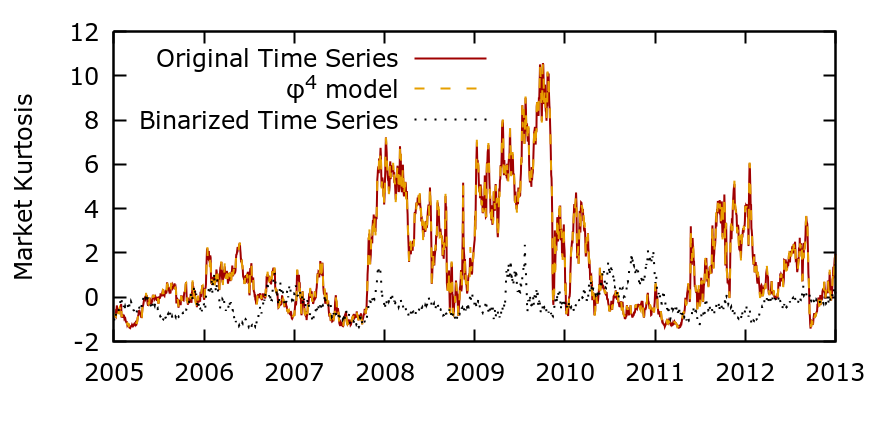}
\caption{\label{fig:fig2} Market mean (top) and market kurtosis (bottom), using a simple moving average of $250$ trading days, versus the date.  The binarized mean value is normalized to reside between the minimum and maximum values of the original time series. } 
\end{figure*}

In this manuscript we map the price changes of stocks to the fields  $\lbrace \phi_{i} \rbrace$ of a $\phi^{4}$ theory. A field $\phi_{i}$ is considered a random variable whose expectation value represents the price change of a stock $i$ on a particular day. We consider a different field theory and set of fields $\lbrace\phi_i\rbrace$ for each point in time and the time evolution of the financial system is represented by a time evolution in model space where the parameters of the theories evolve in time. 

Our aim is to learn the most accurate set of couplings $\theta$ which encode interpretable dependencies among the considered stocks. It is therefore necessary to clarify the role of the inhomogeneous couplings $\theta$ using knowledge of the $Z_{2}$ symmetry. 

Specifically, the inhomogeneous edges $w_{ij}$ act as a weight and are related to the correlation between the two fields $\phi_{i}$ and $\phi_{j}$ \footnote{To give some intuition about how these $w_{ij}$ differ from correlations, one example of their difference is that a correlation between $i$ and $j$ would be absolute, whereas the value of $w_{ij}$ is dynamic and will change depending on what set of stocks the model is learned over.}. When $w_{ij}>0$ two fields $\phi_{i}$ and $\phi_{j}$ are positively correlated, giving rise to a ferromagnetic interaction, whereas when $w_{ij}<0$ they are negatively correlated and are representative of an antiferromagnetic interaction. In addition, the inhomogeneous coupling $a_{i}$ is an external field acting as a bias, which breaks explicitly and locally the symmetry of the system, and biases, depending on its sign, a given field $\phi_{i}$ towards a positive or negative value.  Finally the inhomogeneous couplings $\mu_{i}$ and $\lambda_{i}$ can be tuned to drive a given field $\phi_{i}$ towards either the symmetric or the broken-symmetry phases.

By learning the aforementioned couplings, we are able to extract both internal information, through the set of weights $\lbrace w_{ij} \rbrace$ which express how the price changes of stocks are correlated with each other, and external information, through the set of biases $\lbrace a_{i} \rbrace$ which could express for example how the values of stocks are influenced by external news. We are additionally able to obtain insights into the scaling properties of the $\phi^{4}$ model, by investigating how the weights and biases evolve as we increase the lattice volume of the system or, equivalently, increase the number of modeled stocks.

To solve the problem of learning the most accurate values of the couplings $\theta$ which accurately represent the coexistence and interaction of a number of stocks, we consider the Kullback-Leibler divergence:
\begin{equation} \label{eq:klopp}
KL(q || p) = \int_{-\infty}^{\infty} {q(\bm{\phi})} \ln \frac{q(\bm{\phi})}{p(\bm{\phi}; \theta)} d\bm{\phi}  \geq 0,
\end{equation}
between the Boltzmann probability distribution $p(\bm{\phi},\theta)$ of the $\phi^{4}$ theory and the empirical probability distribution $q(\bm{\phi})$. The empirical probability distribution $q(\bm{\phi})$ in this case simply comprises the values of the empirical time series at a given point in time.

By expanding the terms we obtain:
\begin{equation}\label{eq:klimg}
KL(q || p)= \langle \ln q(\phi) \rangle_{q(\phi)}- \langle \ln p(\phi;\theta) \rangle_{q(\phi)},
\end{equation} 
where $\langle \rangle_{q(\phi)}$ denotes an expectation value under the probability distribution $q(\phi)$. The term $\langle \ln q(\phi) \rangle_{q(\phi)}$ is constant and the minimization of the  Kullback-Leibler divergence is therefore equivalent to the maximization of the remaining term: 
\begin{equation}
\frac{\partial \ln p(\phi ; \theta)}{\partial \theta} = \Big\langle \frac{\partial S}{\partial \theta}\Big\rangle_{p(\phi ; \theta)} -   \Big\langle \frac{\partial S}{\partial \theta} \Big\rangle_{q(\bm{\phi})}.
\end{equation}

The optimization approach is therefore analogous to a maximum log-likelihood estimation, or equivalently, a minimization of the negative log-likelihood. The loss function is therefore:
\begin{equation}
\mathcal{L}=-\frac{\partial \ln p(\phi ; \theta^{(t)})}{\partial \theta^{(t)}},
\end{equation}
and we update the inhomogeneous couplings based on a gradient-based approach as:

\begin{equation}\label{eq:gas}
\theta^{(t+1)}=\theta^{(t)}-\eta  \mathcal{L}.
\end{equation}

Once the optimization process is finished, we have learned the most accurate values of the inhomogeneous couplings within the $\phi^{4}$ theory that are able to reproduce the empirical data as configurations in the equilibrium distribution.

\section{Main results}
\subsection{\label{sec:level3}  Statistics of financial time series}

To illustrate applications of the $\phi^{4}$ scalar field theory for empirical financial data, we consider the log-returns for the time series of a subset of $V=20$ stocks from the S$\&$P 500 index. We assume that the subset of stocks forms a market and calculate statistics, such as the market mean and kurtosis. We consider a time period of $2 \times 10^{3}$ trading days ending on 01/01/2013. The selected time range includes various major financial events, such as the global financial crisis of 2008. To smooth out short-term fluctuations, we apply a simple moving average of $250$ days to the data, which corresponds to one trading year.  Details and definitions are provided in the Appendices.

To establish that the $\phi^{4}$ scalar field theory provides advantages to the modeling of empirical time series in comparison to the binary Ising model, we compare statistics from the true empirical data, the $\phi^{4}$ model, and a binarized version of the original time series. The results are depicted in Fig.~\ref{fig:fig2}. We observe that the $\phi^{4}$ model can faithfully reproduce high-order statistics during major events such as the global financial crisis of 2008. These statistics include the market kurtosis, which can serve as an indicator of financial crises. The results from the $\phi^{4}$ model can be directly contrasted to the binarized version of the original time series that the Ising model would be able to reproduce, where information about high-order statistics such as the kurtosis is completely lost.

\subsection{\label{sec:level3b}  Interpreting the $\phi^{4}$ parameters}

We have previously discussed the conceptual interpretation of the weights $w_{ij}$ within the $\phi^{4}$ theory as a measure capable of capturing correlations between two stocks $i$ and $j$. Here, we investigate whether we can verify numerically that the weights learned by the $\phi^{4}$ theory can accurately reproduce correlations directly observed in experimental data. Specifically, we consider the returns of $20$ stocks from the S\&P 500 index, which are listed in the Appendices, for the date 10/05/2013. We then calculate directly on the data a form of correlation $\rho_{ij}$ between the stocks $i$, $j$ by multiplying the sign of returns $\rho_{ij}=sgn(r_{i})sgn(r_{j})$. If the price of both stocks $i, j$ has increased or decreased on a given trading day, then $\rho_{ij}=1$ and the two stocks are positively correlated. In contrast, if the price of one stock has increased but the price of another stock has decreased on a given trading day then $\rho_{ij}=-1$ and the two stocks are negatively correlated. We then compare these results $\rho_{ij}$ directly calculated from the empirical data against the sign of the weights $sgn(w_{ij})$ learned from the $\phi^{4}$ theory which relate stocks $i$ and $j$.

\begin{figure}[pos=ht]
\center
\includegraphics[width=16cm]{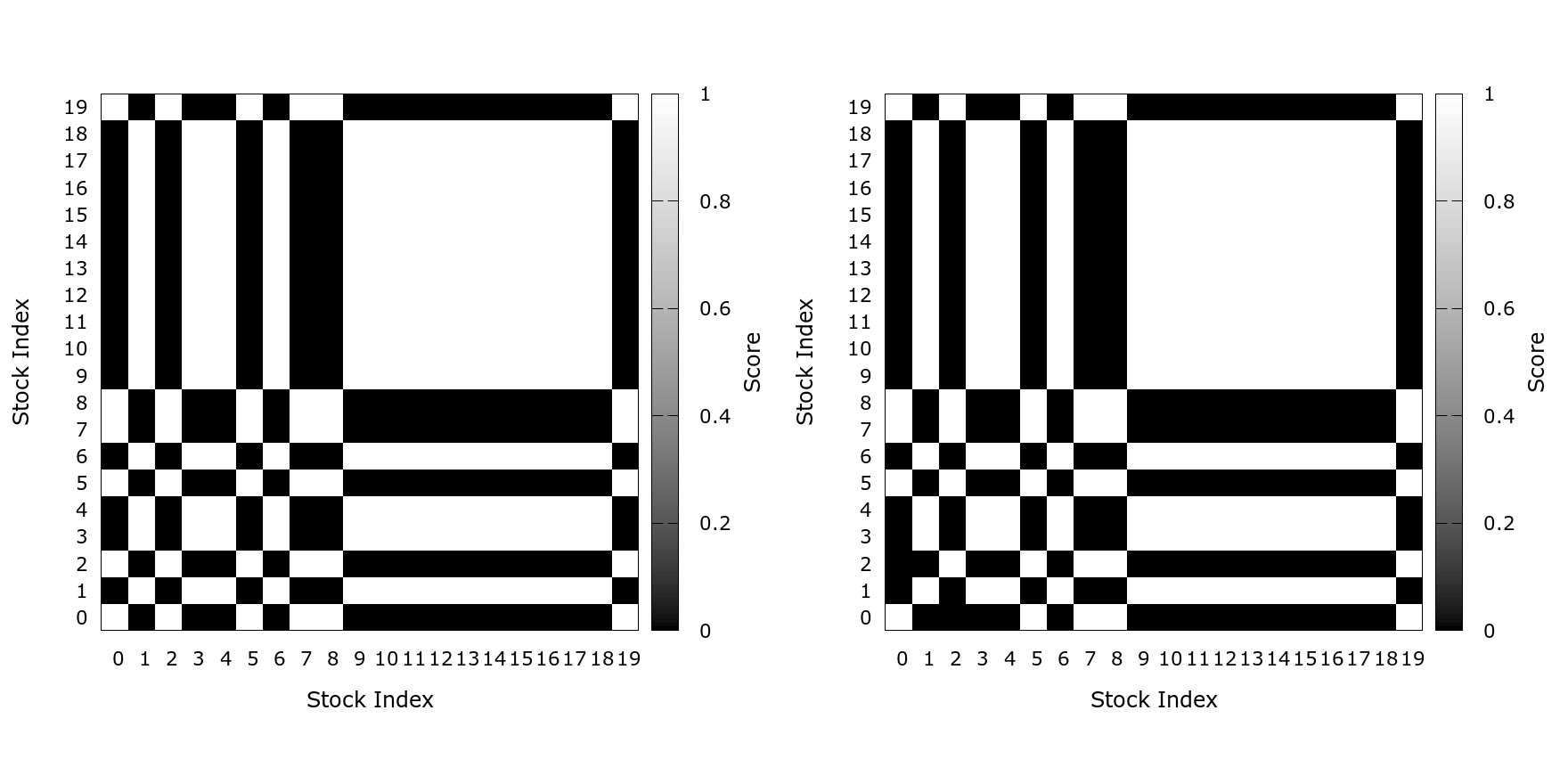}
\caption{\label{fig:corr} The correlations $\rho_{ij}$ between stocks $i$, $j$ directly calculated from experimental data  (left) and the sign of the weights $sgn(w_{ij})$ learned by the $\phi^{4}$ theory (right). } 
\end{figure}

The results are depicted in Fig.~\ref{fig:corr}. We observe an agreement between the true correlations $\rho_{ij}$ and the ones learned from the $\phi^{4}$ theory as depicted via the sign of the weights  $sgn(w_{ij})$. For the training time period considered here, there is only one discrepancy in the learned weights, namely $w_{02}$ (equal to $w_{20}$), which corresponds to the correlation between stocks ABT and AXP. The results support the view that the $\phi^{4}$ theory can properly capture inter-dependencies between stocks.

\subsection{\label{sec:level4} Scaling properties of financial time series}

The set of weights and biases of the $\phi^{4}$ scalar field theory are fully interpretable as they encode either direct correlations between the stocks or correspond to external factors (e.g. news). This is in contrast to machine learning algorithms which utilize hidden variables, where insights into interpretability are more intricate to obtain. Consequently, one is able to increase the lattice volume of the system by increasing the number of stocks modeled, in order to investigate scaling properties of the interpretable weights and biases. This implies that one can extract specific information about each market index in the form of exponents that capture properties of the stocks included within it.

 To investigate scaling properties we then calculate the mean value of weights $\langle w_{ij} \rangle = \frac{1}{N_{w_{ij}}}\sum_{ij}w_{ij}$, and biases $\langle a_{i} \rangle = \frac{1}{N_{a_{i}}}\sum_{i}a_{i}$, where $N$ denotes the number of parameters. Since the weights represent dependencies among stocks, the mean value expresses whether a collection of stocks is positively or negatively correlated. The biases, or external fields, break explicitly the $Z_{2}$ symmetry and their average therefore represents whether a collection of stocks is expected to increase or decrease in value. We then define two exponents $k$ in relation to the number of stocks $V$, or equivalently, the lattice volume $V$ of the system as:
\begin{equation}
\langle w_{ij} \rangle \propto V^{k_{w_{ij}}},\ \langle a_{i} \rangle \propto V^{k_{a_{i}}}.
\end{equation}

We select $V=64$ stocks from the S$\&$P 500 index and calculate the average of their returns over $20$ years at a fixed point in time: 01/01/2023. By considering a large time range, we smooth out temporary local fluctuations. Details about the selection of subsets of $V'=48,32,16$ stocks in order to conduct the scaling calculation are provided in the Appendices.

\begin{figure}[pos=ht]
\center
\includegraphics[width=10cm]{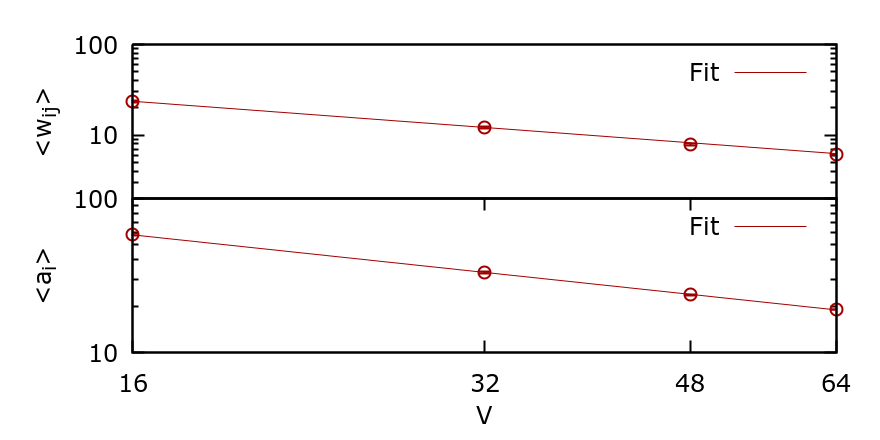}
\caption{\label{fig:fig3} Mean values of the weights (top) and biases (bottom) versus the lattice volume $V$ of the $\phi^{4}$ theory or, equivalently, the number of modeled stocks $V$ from the S$\&$P 500 index. The axes are logarithmic.} 
\end{figure}

The results of the finite size scaling analysis are depicted on logarithmic axes in Fig.~\ref{fig:fig3}. We observe the emergence of scaling properties for the weights and the biases of the $\phi^{4}$ scalar field theory. Specifically, through numerical fits we extract the values of two exponents $k_{w_{ij}}=-0.96(1)$ and $k_{a_{i}}=-0.81(1)$. We repeat the calculation for a different date, namely 01/01/2013, and obtain values of $k_{w_{ij}}=-1.11(4)$ and $k_{a_{i}}=-0.87(3)$.  We remark that analogous calculations of scaling exponents have been conducted before using binarized series and the Ising model~\citep{Borysov2015}. 

The results suggest the presence of nontrivial structure in the weights and biases for the considered S$\&$P 500 index. It is interesting that the results for the two dates are so similar, and warrants further investigation into how the exponents vary over different time periods. Our initial consideration of a large moving average of $20$ years was to investigate the asymptotic limit, but it may be that the scaling properties for shorter time periods could be exploited in investment strategies. 

\subsection{\label{sec:level51} Predicting missing values in financial time series}

 Before illustrating the forecasting capabilities of the $\phi^{4}$ theory we first consider an easier problem of filling in missing data. We train the model on a dataset comprising the returns for the time series of the AAPL, MSFT and NVDA stocks. We then employ techniques of probabilistic machine learning, pertinent to the sampling of missing degrees of freedom, to predict the price changes of stocks for trading days that are not included in the training dataset. 

In more detail, the question we aim to answer for a specific trading day is ``If the price change of the AAPL stock is $\phi_{AAPL}$, and the price change of the MSFT stock is $\phi_{MSFT}$, what will the price change $\phi_{NVDA}$ of the NVDA stock be?". In a mathematically formal representation, this question can be expressed in terms of conditional probability distributions as:
\begin{equation}
p(\phi_{NVDA} | \phi_{AAPL},\phi_{MSFT}).
\end{equation}

\begin{figure}[pos=ht]
\center
\includegraphics[width=10cm]{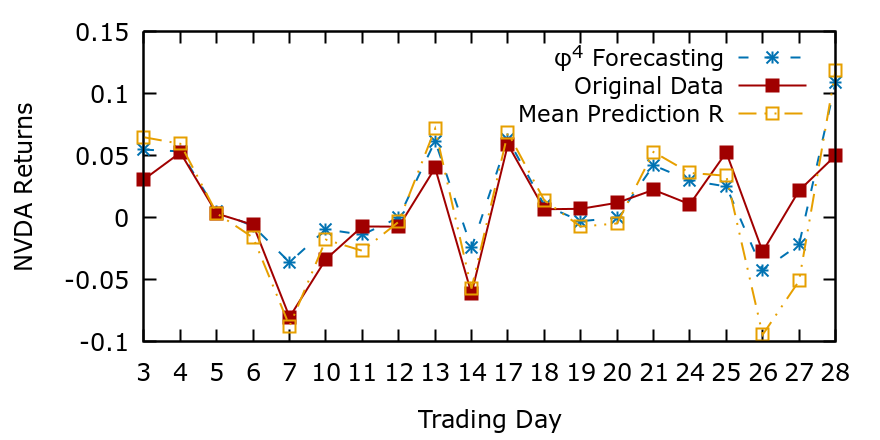}
\caption{\label{fig:fig4} NVDA returns versus the trading day, for the original time series, predictions from the $\phi^{4}$ model and a rescaled mean calculation using the simultaneous returns of AAPL and MSFT as predictors. The x axis corresponds to dates for October 2022. } 
\end{figure}

The results, obtained by sampling the conditional probability distribution, are depicted in Fig.~\ref{fig:fig4}. We compare the $\phi^{4}$ result for the NVDA stock against a baseline prediction $R$ calculated based on the rescaled mean of the two remnant stocks as:
\begin{equation}
R_{NVDA}= \frac{\sigma_{NVDA}}{2}\Bigg( \frac{\phi_{AAPL}}{\sigma_{AAPL}}+\frac{\phi_{MSFT}}{\sigma_{MSFT}} \Bigg),
\end{equation}
where $\sigma$ corresponds to the standard deviation of a stock in the time window of the training dataset.

We observe approximate agreement with the true data although with inevitable discrepancies since the price change of the stock will also be influenced by other factors, for example momentary news that the algorithm currently has no access to. The mean absolute errors for the $\phi^{4}$ result and the rescaled mean prediction are $0.019$ and $0.023$, respectively. The $\phi^{4}$ result performs slightly better, but over too little data to make a firm statement and for the moment we consider the two methods to be of comparable performance. However, importantly, the rescaled mean method can only be employed in cases of highly correlated stocks, whereas we can expect the $\phi^{4}$ method to also be applicable to situations with lower correlations.  The above calculation in its current form cannot be exploited for trading, as the opening times of the predictor and predicted stocks are simultaneous. In order to convert it to a forecasting setup, it could be used on different correlated stocks with disjoint opening times, or by shifting the predicted stock's returns forward by one time step in the training stage. 

\subsection{\label{sec:level52} Forecasting financial time series}

A simpler forecasting set up that can be potentially utilized for trading is to forecast a stock price for the upcoming trading day, based on that stock's historical returns. To obtain a prediction for the upcoming trading day with the $\phi^{4}$ scalar field theory, we consider the case of the AAPL stock, and construct a dataset where each training point comprises the returns $\phi_{i}$ of a given trading day $i$ and the preceding 149 trading days.  If we denote future and past trading days with positive or negative subscripts, respectively, and the returns of the current trading day as $\phi_{0}$, the $\phi^{4}$ machine learning algorithm learns a joint probability distribution $p(\lbrace\phi_{i}, \phi_{i-1},\ldots, \phi_{i-149}\rbrace)$, where $i\leq 0$. One then observes that once the closing price $\phi_{0}$ of a stock is obtained for the current trading day one can utilize the learned weights and biases encoded in the aforementioned joint probability distribution to ask the question  ``What will the price change $\phi_{1}$ of the AAPL stock be tomorrow, given the price change $\phi_{0}$ today and the price changes $\phi_i$ of the past 148 trading days?", or formally, one can sample the conditional probability distribution:
\begin{equation}
p(\phi_{1} | \phi_{0},\ldots,\phi_{-148}).
\end{equation}

\begin{figure}[pos=!ht]
\center
\includegraphics[width=10cm]{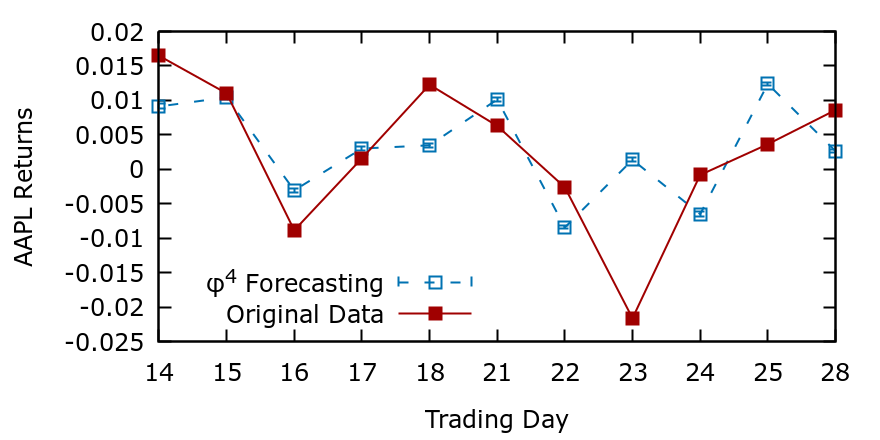}
\caption{\label{fig:fig5} Forecasting of the AAPL returns for the upcoming trading day using the $\phi^{4}$ theory with historical AAPL returns as predictor. The x axis corresponds to dates for October 2024.} 
\end{figure}

An example of predicting the closing price for the upcoming trading day in October 2024 for the case of the AAPL stock is depicted in Fig.~\ref{fig:fig5}. 

To check the accuracy of the prediction we run incremental forward validation and statistical significance testing over multiple time periods. Specifically, we consider a time period of three months, extending from the start of October to the end of December 2024, and train six $\phi^{4}$ models, two per month, to forecast the closing price for the upcoming trading days. Each model predicts a time period of two weeks. In order to compare our $\phi^{4}$ algorithm to reasonable baseline algorithms, we also calculate the results for an LSTM with an identical data set up and a simple linear regression with a rolling window. Details of these two algorithms are provided in the Appendices. 

\begin{figure}[pos=!ht]
\center
\includegraphics[width=10cm]{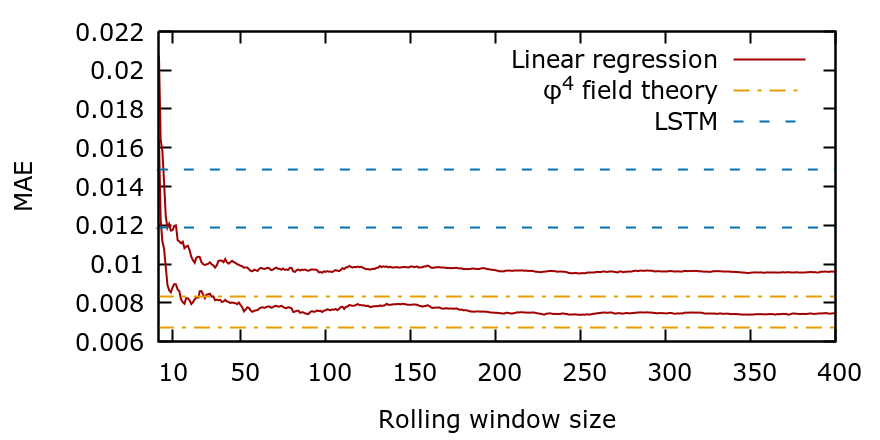}
\caption{\label{fig:fig6}  Mean absolute error for the forecasting of AAPL price changes using linear regression versus the rolling window size, and comparison with the $\phi^{4}$ theory and an LSTM. The statistical uncertainty is shown by the gap between the two lines depicted for each algorithm.}
\end{figure}

We select as a metric the mean absolute error. The comparison of the three algorithms for the three month period is depicted in Fig.~\ref{fig:fig6}.  Modeling the financial time series with $\phi^{4}$ quantum field theory results in a lower value of the mean absolute error in comparison to the LSTM and linear regression for all window sizes of the linear regression up to $400$ days. The effective training window of the $\phi^{4}$ model and the LSTM was chosen to be $230$ days.
We remark that the LSTM has the highest mean absolute error and thus the lowest performance from among the three approaches. We attribute the lower performance of the LSTM to the fact that the data has a very high noise to signal ratio so that the LSTM very easily overfits when it does not have sufficient training data.

An additional question in relation to the comparison of the three algorithms is that of computational complexity. We remark that the $\phi^{4}$ model has analogous computational complexity to other statistical-mechanical or energy-based models, for example the Ising model, since it necessitates the implementation of a Markov chain Monte Carlo simulation with the Metropolis algorithm. This implies that the $\phi^{4}$ model is computationally demanding in comparison to a linear regression but the results indicate that it can possess lasting predictive power that can extend for time periods of two weeks. LSTMs are neural networks with hidden variables, the compute for training scales linearly with number of network parameters and number of tokens in the dataset and we note that the hyperparameter tuned LSTM used in this work has around $13 \times 10^{3}$   model parameters.

\section{\label{sec:level6} Discussion and conclusions}

We proposed a disordered $\phi^{4}$ quantum field theory, which is mathematically equivalent to a machine learning algorithm of a Markov random field, to model and forecast financial time series. 

Specifically, we have shown advantages of the $\phi^{4}$ quantum field-theoretic machine learning algorithm in relation to the Ising model by demonstrating that it can reproduce high-order statistics for empirical financial time series, such as the market kurtosis, which can serve as an indicator of  financial crises. This crucial information is lost in binarized series reproduced by Ising models. We additionally investigated scaling properties for the S$\&$P 500 index by extracting exponents which govern the behavior of the learned weights and biases in relation to the number of stocks. Finally, we employed the $\phi^{4}$ scalar field theory to forecast price changes of AAPL, MSFT, and NVDA stocks.

Our initial results show some promise that the $\phi^{4}$ model could be used to build investment strategies, for instance by filling in missing data, predicting the price change of a stock in a market with a disjoint opening time or forecasting the price change of an upcoming trading day. We have illustrated improvements in performance over some simple baseline strategies. The purpose of this was to demonstrate that this approach has potential, not yet to present fully formulated investment strategies which could be employed in live trading. We leave to future work further exploration of the model's capabilities.  
The scaling results for the weights and biases of the theory hint at the possibility of uncovering features of market structure, which once more fully explored may be able to be exploited. The external biases could also be specified to enrich the behavior the model can capture. Finally, the model may have sufficient predictive power for more than one day ahead prediction and forecasts a full probability distribution, and so could be used to produce predictions for option prices, or in transaction cost minimizing trajectories~\citep{pedersen}.

We remark that one envisages, based on the $\phi^{4}$ model, an opportunity to develop a framework for financial markets within quantum field theory. Specifically, the $\phi^{4}$ theory discussed herein can be viewed as a dimensionally compactified version of a generalized model. Explicitly, one is able to progressively add dimensions to the system, such as a dimension which corresponds to the time evolution of the series, or an additional dimension that relates stocks positioned within market indices of different countries. These generalized and d-dimensional $\phi^{4}$ scalar field theories would then be capable of extracting a set of more sophisticated dependencies. Via a simultaneous application of the mathematical operations of dimensional compactification and renormalization of the learned couplings one would then be able to recover mathematically either the model discussed in this manuscript or one representing intricate dependencies between multiple market indices. 

To summarize, we have shown advantages of the $\phi^{4}$ quantum field theory, a system that is able to directly model the continuous values of price changes for stocks, in relation to spin models traditionally employed in applications of econophysics which necessitate a binarization of time series and lead to loss of crucial information about the structure of financial markets. As a result, we anticipate merits in the implementation, investigation, and extension of ideas from quantum field theory to the research field of finance.

\section*{\label{sec:level11}Data Availability Statement} Code to implement the disordered $\phi^{4}$ machine learning algorithm for financial data will be made publicly available with the published manuscript on ~\footnote{{https://github.com/dbachtis/phi4fin}}. Data which support the results of this manuscript are available from the authors upon request. Experimental data for financial time series are obtained from the Wharton Research Data Services (WRDS), a proprietary, subscription-based data platform, and are not available from the authors upon request.

\section*{\label{sec:level12}Acknowledgements} This research was supported by the Science and Technology Facilities
Council (STFC) Consolidated Grant ST/X00063X/1 ``Amplitudes, Strings \& Duality" and in part by grant no. NSF PHY-2309135 to the Kavli Institute for Theoretical Physics (KITP). This project has not received commercial funding. 


\appendix

\section{The $\phi^{4}$ theory for financial data}

The Ising model has been traditionally employed to study aspects of finance due to the presence of the $Z_{2}$ symmetry. The $Z_{2}$ symmetry allows for the representation of various dynamics, such as the decision of an agent at a given point in time to buy or sell a stock. In the context of financial time series, the value of a stock can either increase or decrease on a given trading day (by a large or a minuscule amount) hence allowing systems with $Z_{2}$ symmetries to emerge as natural candidates for the representation of such problems. As discussed in the main manuscript, empirical financial time series model quantities such as log-returns which are represented by continuous values. To model these quantities with the Ising model, which is only capable of representing binary degrees of freedom, one would need to discretize or binarize the financial time series, for instance by considering the sign of returns. The simplification of financial data necessitated by the Ising model gives rise to inaccuracies in the study of empirical financial time series with spin models.

In this context, the $\phi^{4}$ scalar field theory emerges as a potential $Z_{2}$ symmetric system capable of modeling financial time series. The $\phi^{4}$ scalar field theory is a generalization of the Ising model, and includes a reducible limit under which one is able to mathematically recover the Ising model. This implies that the $\phi^{4}$ scalar field theory should be as successful as the Ising model, with the additional benefit of being able to model continuous degrees of freedom. An additional benefit of the $\phi^{4}$ scalar field theory is that it is a nonlinear system, allowing for the representation of intricate dependencies on data, while also retaining complete interpretability. Other types of continuous quantum field theories are either too simple, such as the free field which extracts linear dependencies and would compete unfavorably against other simple algorithms, or too complicated, hence making their implementation questionable based on the current published literature. We therefore believe the $\phi^{4}$ quantum field theory strikes a balance between complexity, simplicity, interpretability, and representational capacity for the discussed scientific problems.

\begin{figure}[pos=ht]
\center
\includegraphics[width=10cm]{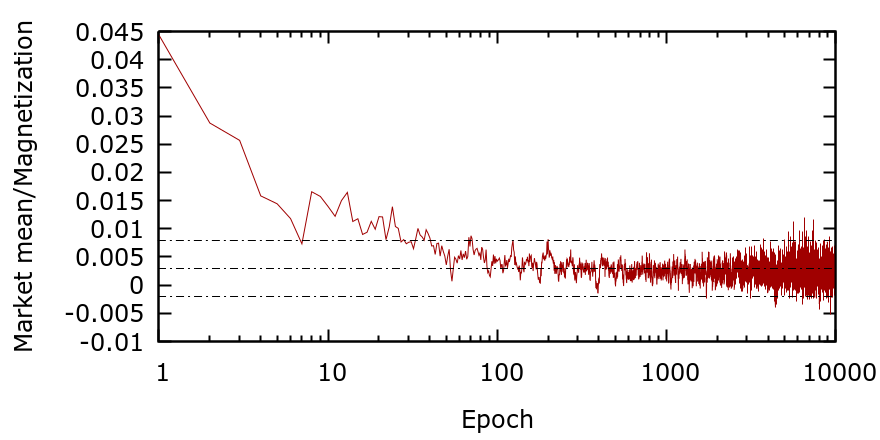}
\includegraphics[width=10cm]{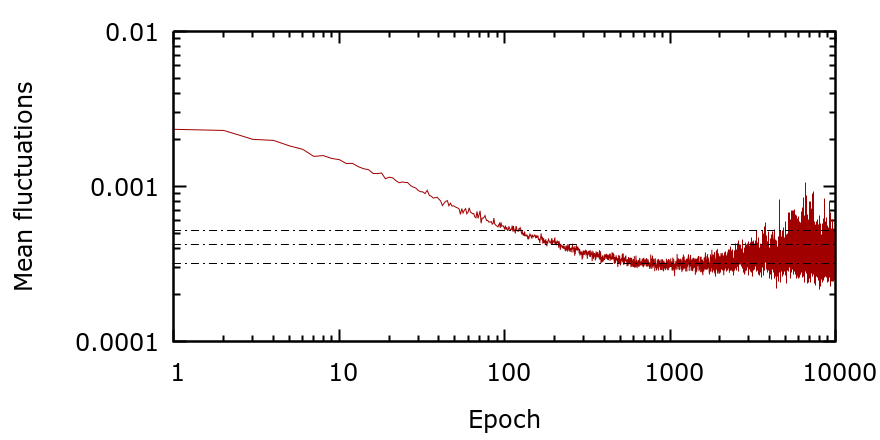}
\caption{\label{fig:train} Expectation value of the market mean/magnetization (top) and of the fluctuations (bottom) versus the number of epochs. The dashed lines correspond to the true values including the statistical uncertainty.} 
\end{figure}

\subsection{Definitions}

 The $\phi^{4}$ scalar field theory is a system described by continuous degrees of freedom, which implies that for each field $\phi_{i}$,  $-\infty< \phi_{i} <\infty$. In this manuscript we are interested in modeling a set of continuous values which correspond to financial data and therefore reside in a given range. Consequently, we need to set bounds in the sampling process of the $\phi^{4}$ theory. 

A normalization which fully preserves the structure of the experimental data is a simple rescaling: 
\begin{equation}
X'=X/abs[max(X)],
\end{equation}
which rescales the largest value of the data in terms of magnitude to be +1 or -1 depending on its sign. Rescaling the data can allow, in principle, for easier training of the $\phi^{4}$ machine learning algorithm depending on how its parameters have been initialized. This can be easily understood by considering the Gaussian case for $\lambda=0$ where it is possible to arrive at a closed form expression in which the standard deviation of the data is $\sigma \propto 1/\sqrt{mu^{2}}$, see Ref.~\citep{bachtis2024latticephi4fieldtheory} for a derivation. This implies within the Gaussian case that for data described by small values of standard deviation, large values of the squared mass are expected to be learned. Once the $\phi^{4}$ machine learning algorithm has been trained, one can always transform the sampled data to the original range via:
\begin{equation}
X=X'abs[max(X)].
\end{equation}

 A different normalization is to rescale the data to reside within the range $[-1,1]$. Before training the $\phi^{4}$ machine learning algorithm we first normalize the complete set of financial data $X$ in the range $[-1,1]$ as:
\begin{equation}
X'=2\frac{X-\min(X)}{\max(X)-\min(X)}-1.
\end{equation}

We remark that the current normalization is not robust to outliers, and we plan to investigate other normalization procedures in future work. To propose a new degree of freedom for the Markov chain Monte Carlo implementation using the Metropolis algorithm we then sample uniformly in the range $[-1.5,1.5]$. Once the training process is completed, we sample a set of data from the $\phi^{4}$ theory and we rescale both the sampled and the transformed data in the range of the original values as:
\begin{equation}
X=\frac{(X'+1) (\max(X)-\min(X))}{2}+\min(X).
\end{equation}

\begin{figure*}[pos=ht]
\center
\includegraphics[width=10cm]{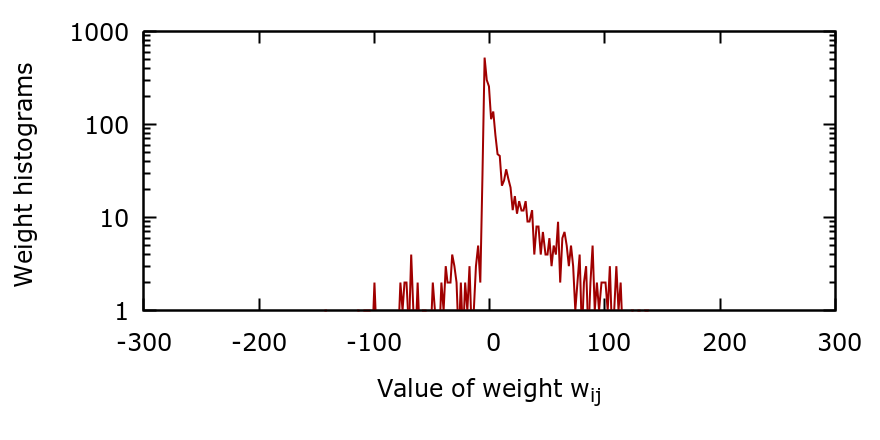}
\includegraphics[width=10cm]{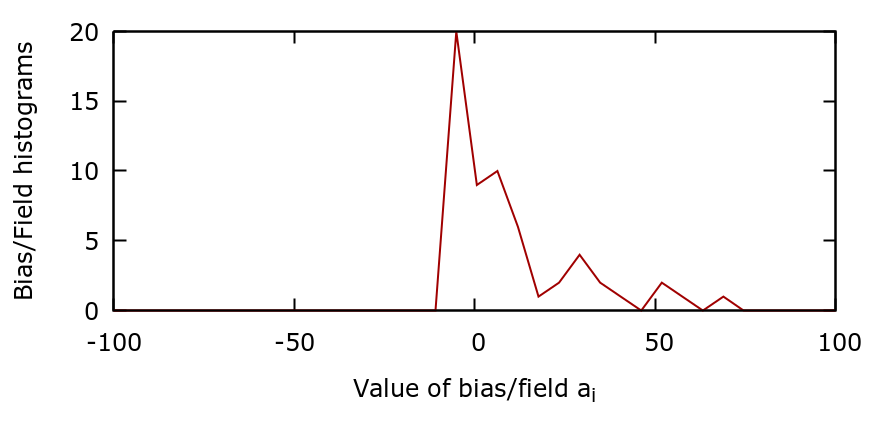}

\caption{\label{fig:dist} Histograms of the weights and biases for a simple moving average of $5 \times 10^{3}$ trading days, and $V=64$ modeled stocks from the S$\&$P 500 index.} 
\end{figure*}
We define a daily closing price as $c(t)$ and the log-returns and returns as:
\begin{equation}
r_{\log}(t)=\ln\Bigg[\frac{c(t)}{c(t-1)}\Bigg],
\end{equation}
\begin{equation}
r(t)=\frac{c(t)-c(t-1)}{c(t-1)}.
\end{equation}

The values of returns used in the manuscript to forecast financial time series are corporate action adjusted. In contrast, the values of log-returns used to model financial time series and reproduce high-order statistics have not been adjusted for corporate actions. 

We define a simple moving average, $SMA$, at day $t$, as a mean over the values of the $n$ previous days including day $t$:
\begin{equation}
SMA(t)=\frac{\sum_{i=0}^{n-1} r(t-i)}{n}.
\end{equation}

\subsection{\label{sec:level8}  Scaling properties of the weights and biases}

Here, we provide further details about the calculation of exponents for the weights and biases of the $\phi^{4}$ machine learning algorithm. A crucial remark  is that for the study of scaling properties of Fig.~3 in the main text, we consider the values of the squared mass and the lambda couplings as global, namely $\lambda_{i}=\lambda$ and $\mu^{2}_{i}=\mu^{2}$. This allows one to keep the squared mass and lambda couplings fixed while the lattice volume of the system is varied and to extract properties for the remaining inhomogeneous couplings. To learn a set of global couplings we first train the $\phi^{4}$ model for a given lattice volume, such as $V=64$. We then fix the squared mass and lambda couplings with the learned global values and train to learn the remnant inhomogeneous and local couplings for the smaller lattice volumes of $V'=48,32,16$.  For the calculation of mean values in subsets of $V'=48,32,16$ stocks, we randomly select, without replacement, $V'$ stocks from the complete set of $V=64$. To diminish errors from the specific selection of the randomly selected $V'$-size subsets we repeat the process for a number of iterations equal to $40,40$ and $100$, respectively. 

We remark that there are various sources of systematic and statistical uncertainties included in the numerical fits. One source of uncertainty is introduced by the fact that we select randomly the subset of stocks in order to conduct the finite-size scaling analysis, which implies that one needs to increase the number of iterations in this random selection, especially as the size of stocks is reduced. Another source of uncertainty is introduced by the training process of the algorithm. This second source of uncertainty is reduced when a distinct $\phi^{4}$ machine learning algorithm is trained for each iteration, and is taken into consideration in the manuscript. Another source of uncertainty relates to the fact that the numerical fit calculates a scaling exponent which necessitates an extrapolation to infinity. Accurate calculation of the exponents hence requires extensive computational calculations.

\begin{figure}[pos=ht]
\center
\includegraphics[width=10cm]{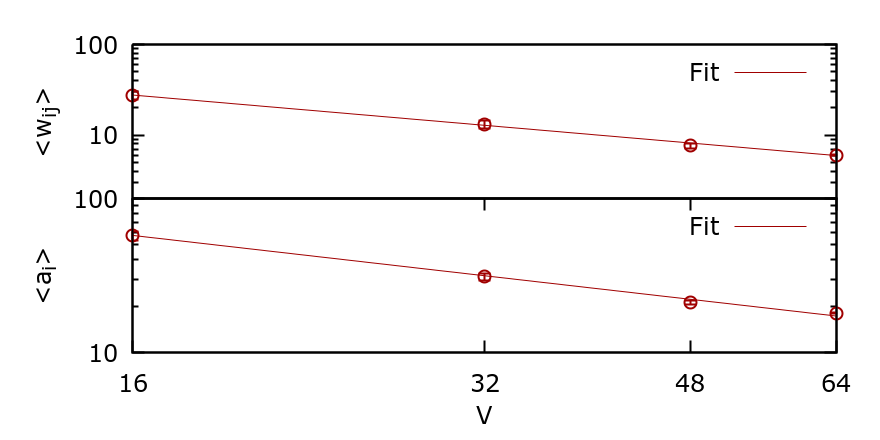}

\caption{\label{fig:expo} Calculation of the exponents for the weights (top) and the biases (bottom) with a moving average of $5000$ trading days and a starting date of 01/01/2013. } 
\end{figure}

In the current numerical fits within the manuscript, which consider a moving average of $5000$ trading days, which corresponds to $20$ trading years, the RMS of residuals for the calculation of the exponent of the weights is $0.78$ and for the one of the biases is $0.84$. We additionally conduct another calculation of exponents with a starting date of 01/01/2013 and using a random selection of $40,40,50$ subsets for the $48,32,16$ stocks. The results are depicted in Fig.~\ref{fig:expo}, where we obtain the exponents $k_{w_{ij}}=-1.11(4)$ and $k_{a_{i}}=-0.87(3)$ with RMS of residuals equal to $0.47$ and  $0.86$, respectively.

We additionally provide the supporting Fig.~\ref{fig:dist}, which depicts the histograms for the weights and biases in the case of a moving average of $5 \times 10^{3}$ and for $V=64$ stocks. The time average corresponds to the setting of the finite-size scaling analysis employed in the main manuscript in order to calculate exponents. From the distribution of the couplings we observe the anticipated behavior, related to the simple observation that for very large timeframes, such as the time span of $20$ trading years considered here, financial markets will generate profits which implies that one expects a higher representation of positive couplings. See the main text for a description of the information captured by the interpretable weights and biases.

\subsection{\label{sec:level9}  Predicting price changes}
In the implementation of the $\phi^{4}$ machine learning algorithm to forecast price changes or to sample the returns of correlated stocks we separate the time series in a training and prediction set. The training set comprises the chronologically former values and the prediction set the most recent ones. In the figures we depict the dates for which the prediction has been obtained. The training dataset precedes these days. For the figures relating to the sampling of correlated stocks we consider a dataset of 230 preceding days. In the context of forecasting the AAPL stock each data point comprises the values of returns from the past 150 trading days. For example, the first data point includes the values $t_{-10}, t_{-11},\ldots, t_{-159}$, and the second data point includes the values $t_{-11},t_{-12},\ldots,t_{-160}$. For forecasting the price change in the upcoming trading day of the AAPL stock we consider  80 data points, hence resulting in an effective window of $230$ trading days.  

\begin{figure}[pos=ht]
\center
\includegraphics[width=10cm]{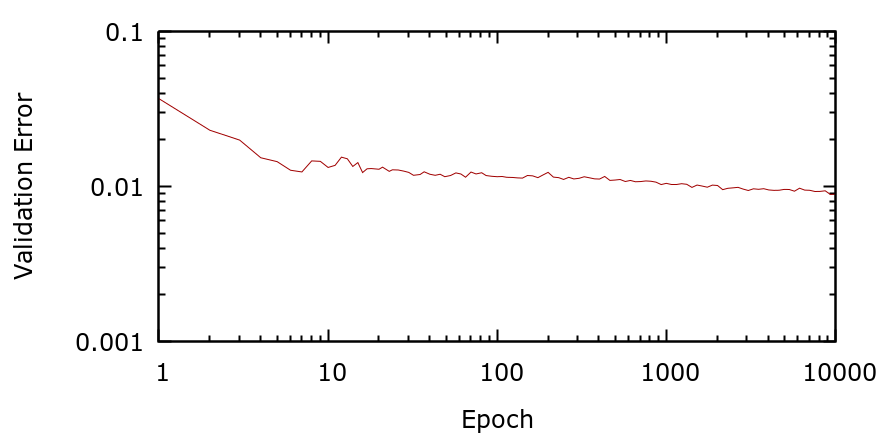}
\includegraphics[width=10cm]{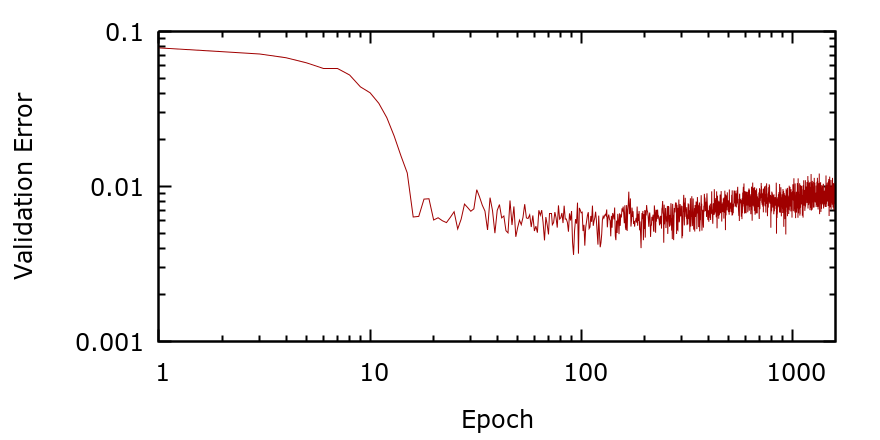}

\caption{\label{fig:valerr} Validation errors for the case of three correlated stocks (top) and for the forecasting of the AAPL stock (bottom). } 
\end{figure}

We additionally monitor the accuracy of the machine learning algorithm using a validation set which is a subset of the training set. Specifically, for both the cases of predicting the returns of correlated stocks and forecasting the returns of the AAPL stock we construct a validation set by excluding $10$ randomly selected data points from the training process. In the case of correlated stocks the validation error corresponds to the average of the mean absolute errors between the predicted and true values for each of the three stocks. In the case of forecasting the AAPL stock the validation error is the mean absolute error between the predicted and true value.  The validation errors are depicted for both cases on Fig.~\ref{fig:valerr}. We avoid overfitting on the data via the process of early stopping.

For the learning of the probability distribution in the context of sampling the correlated stocks, we observe that the market mean $MM$, defined as an average over the values of stocks $r_{i}$:
\begin{equation}
MM=\frac{1}{V} \sum_{i} r_{i},
\end{equation}
is mathematically equivalent to the calculation of the intensive magnetization $m$ for the $\phi^{4}$ scalar field theory, namely:
\begin{equation}
m=\frac{1}{V} \sum_{i} \phi_{i},
\end{equation}
since the returns of the stocks $r_{i}$ have been directly mapped to the degrees of freedom $\phi_{i}$ of the $\phi^{4}$ theory. Consequently, to monitor the learning process of the $\phi^{4}$ scalar field theory in relation to the empirical probability distribution represented by the dataset we observe the evolution of the expectation value of the magnetization $\langle m \rangle$ and its fluctuations $\chi_{m}=\langle m^{2} \rangle-\langle m \rangle^{2}$. The results are depicted on Fig.~\ref{fig:train} where they are directly compared with the true values.

In the main manuscript, we studied the case of fixing the values of two stocks and sampling the remaining one. It is also possible to investigate the case of fixing only one of the stocks and sampling the remaining two. In this setting the scientific question asked is ``If the price change of the AAPL stock is $\phi_{AAPL}$, what will the price changes $\phi_{NVDA}$, $\phi_{MSFT}$ of the NVDA and MSFT stocks be?". This is represented by the conditional probability distribution:
\begin{equation}
p(\phi_{NVDA}, \phi_{MSFT} | \phi_{AAPL}).
\end{equation}

\begin{figure*}[pos=ht]
\center
\includegraphics[width=10cm]{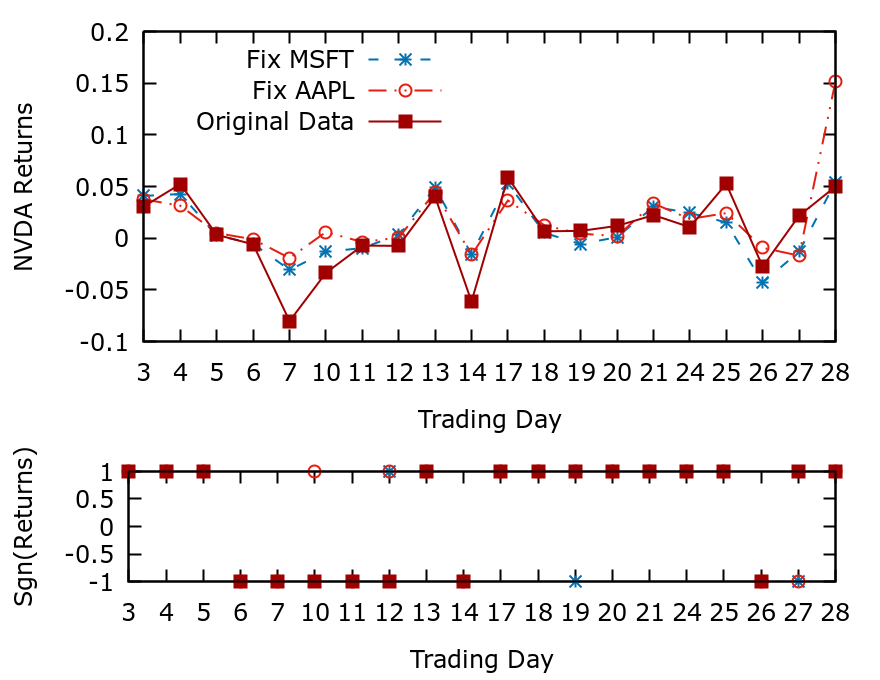}

\caption{\label{fig:f5} NVDA returns versus trading days for October 2022 in the context of fixing only one of the values of the MSFT or AAPL stocks .} 
\end{figure*}

\begin{figure*}[pos=ht!]
\center
\includegraphics[width=8cm]{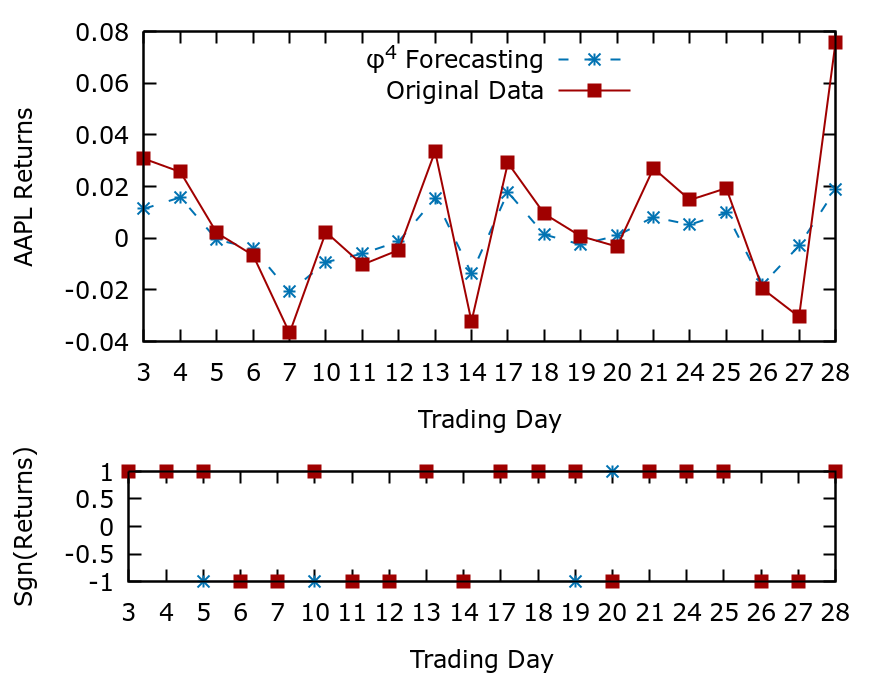}
\includegraphics[width=8cm]{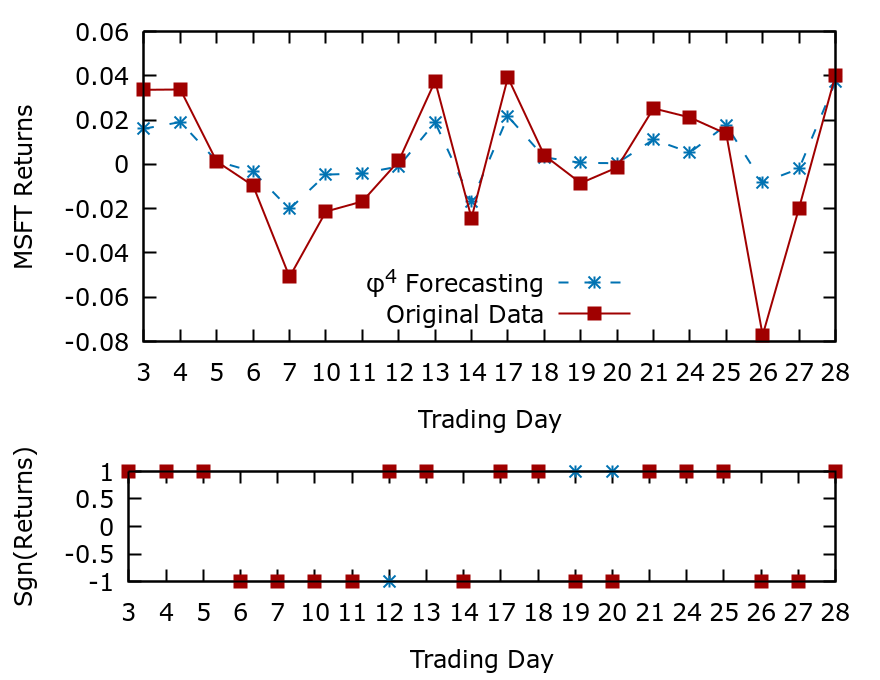}
\includegraphics[width=8cm]{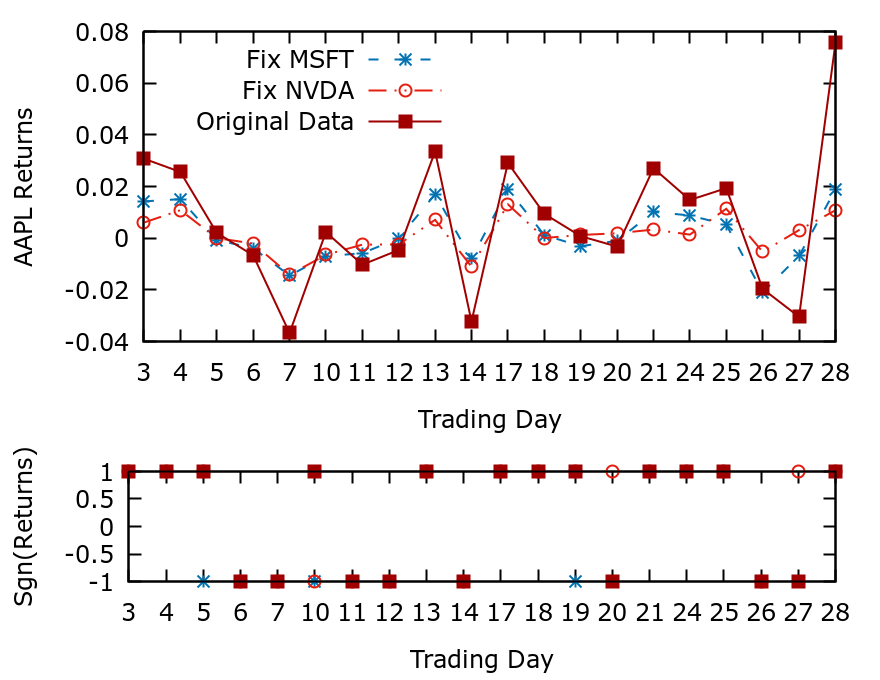}
\includegraphics[width=8cm]{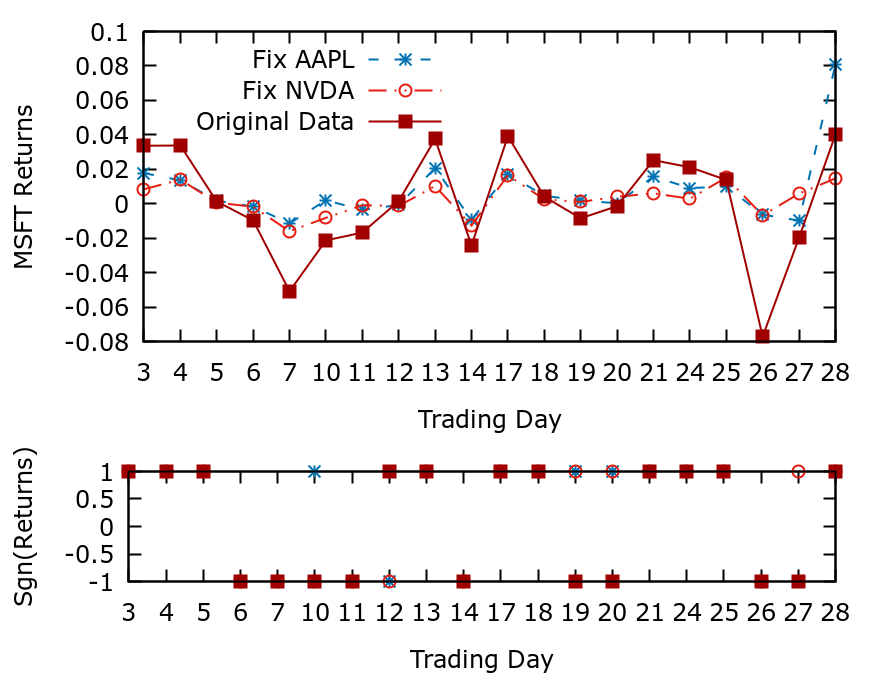}

\caption{\label{fig:fore} Forecasting of time series for returns of stocks versus trading days for October 2022.} 
\end{figure*}

 This is an intricate scientific question which assumes that the price changes of both the NVDA and MSFT stocks are heavily, if not exclusively, dependent on the price change of the AAPL stock. The results are depicted in Fig.~\ref{fig:f5}, where we observe that some information about the price change is still preserved in the predictions. We clarify that the aforementioned method is only applicable when the correlations are strong, which implies that the values of the weights $w_{ij}$ are significant. In Fig.~\ref{fig:fore} we depict the analogous figures pertinent to the sampling of correlated stocks for the two remaining cases of AAPL and MSFT.

We remark that in the comparison of the $\phi^{4}$ quantum field theory against the linear regression, the linear regression is trained on all data up to the previous trading day, while, in contrast, the $\phi^{4}$ model has only been trained up to the start of a $10$ trading day window, and then the most recent data fed through the trained model to predict the future price change of the AAPL stock. Given this difference in training data, the fact that the $\phi^4$ model still performs better than the linear regression indicates that the $\phi^{4}$ system may have more multi-day predictive power. The $\phi^{4}$ model could also be retrained every day with the closing price of a stock to produce more competitive results than the ones depicted in the manuscript.

 We trained a Long Short Term Memory (LSTM) time series model on the same datasets we used for the $\phi^4$ model. The model was implemented in PyTorch using the standard LSTM library package. In addition to a vanilla LSTM we also included a dropout layer and incorporated weight decay with the adamW optimiser. The following hyperparameters of the LSTM (and its training) were investigated using the Bayesian optimisation package {\it{optuna}}:  hidden dimension,  number of layers, dropout, learning rate, batch size, weight decay. The result of the hyperparameter optimisation produced the following choice for these parameters:
Hidden dimension=32; number of layers=2; dropout=0.3; learning rate= 0.01; batch size= 32; weight decay=0.0001. The LSTM was run for 500 epochs. The results showed larger MAE loss than both regression and the $\phi^4$ model. This is not necessarily surprising,  although LSTMs are capable of representing long-range nonlinear temporal dependencies, their representational capacity doesn't imply superior forecasting performance for financial time series. Systematic evaluations have found that LSTMs frequently fail to outperform naïve or conventional statistical benchmarks when training samples are limited~\citep{Buczynski2023,Kobiela2022,Gul2026}. Moreover, even strong historical results may be unstable over time: \citep{FischerKrauss2018}. This indicates the high noise to signal ratio and non-stationarity of the dataset.

\section{\label{sec:level10} Interpretation of the $\phi^{4}$ parameters}

To obtain an intuitive understanding of the types of dependencies extracted via the machine learning implementation on the set of financial data, we observe the update rules of the parameters $\theta$, listed in the main manuscript, and the fact that they depend on a derivative of the lattice action $S$ versus the parameter $\theta$ as ${\partial S}/{\partial \theta}$ . For the case of the weights the update rule then depends on the term $\phi_{i}\phi_{j}$ that extracts a coupling, which is related to the correlation between the fields $\phi_{i}$ and $\phi_{j}$ which correspond to the values of returns of two distinct stocks $i$ and $j$. The parameters $w_{ij}$ have therefore the role of an unnormalized weight, which measures the strength of dependency between two stocks. The biases $a_{i}$ then depend directly, via the derivative of the lattice action, on the values of the returns. A bias, depending on its sign, encodes by a given magnitude whether the price of a stock is expected to increase or decrease in value. We recall that, in this lattice and Boltzmann-probabilistic representation, a field $\phi_{i}$ corresponds to a random variable $\phi$ on a site $i$, which represents the i-th stock. The random variable $\phi_{i}$ can then correspond to either the return of a stock at a fixed point in time or the return of a stock in a time window, represented as an expectation value. For the numerical interpretation of the sign of the weights $w_{ij}$ in relation to the true correlations as calculated directly on experimental data we considered 20 stocks, namely "ABT", "AMGN", "AXP", "BAC", "BMY", "CAT", "CL", "COP", "CVX", "DE", "DHR", "EMR", "EXC", "GE", "HD", "HPQ", "INTC", "MSFT", "ORCL", "AAPL".

\newpage
\bibliographystyle{unsrt}
\bibliography{ms2}

\end{document}